\documentclass[conference]{IEEEtran}
\IEEEoverridecommandlockouts
\usepackage{cite}
\usepackage{amsmath,amssymb,amsfonts}
\usepackage{algorithmic}
\usepackage{graphicx}
\usepackage{textcomp}
\usepackage{xcolor}
\usepackage{booktabs}
\def\BibTeX{{\rm B\kern-.05em{\sc i\kern-.025em b}\kern-.08em
    T\kern-.1667em\lower.7ex\hbox{E}\kern-.125emX}}
\begin{document}

\title{CSI-Agnostic End-to-End Neural Decoding of LDPC Codes under Synchronization Errors
}
\author{\IEEEauthorblockN{Yi Kuang}
\IEEEauthorblockA{\textit{Institute of Artificial Intelligence} \\
\textit{CISDI Information Technology(Chongqing) Co., Ltd.}\\
Chongqing, 401122, China \\
schemer@sjtu.edu.cn}
\and
\IEEEauthorblockN{Shu Liu}
\IEEEauthorblockA{\textit{National Key Laboratory of Wireless Communications} \\
\textit{University of Electronic Science and Technology of China}\\
Chengdu, China\\
shuliu@uestc.edu.cn}
\and
\IEEEauthorblockN{Ningning Wang}
\IEEEauthorblockA{\textit{School of Computer Science} \\
\textit{Shanghai Jiao Tong University}\\
Shanghai, China \\
wangnning@sjtu.edu.cn}
\iffalse
\and
\IEEEauthorblockN{4\textsuperscript{th} Given Name Surname}
\IEEEauthorblockA{\textit{dept. name of organization (of Aff.)} \\
\textit{name of organization (of Aff.)}\\
City, Country \\
email address or ORCID}
\and
\IEEEauthorblockN{5\textsuperscript{th} Given Name Surname}
\IEEEauthorblockA{\textit{dept. name of organization (of Aff.)} \\
\textit{name of organization (of Aff.)}\\
City, Country \\
email address or ORCID}
\and
\IEEEauthorblockN{6\textsuperscript{th} Given Name Surname}
\IEEEauthorblockA{\textit{dept. name of organization (of Aff.)} \\
\textit{name of organization (of Aff.)}\\
City, Country \\
email address or ORCID}
\fi
}

\author{
\parbox{0.32\textwidth}{
\centering
Yi Kuang \\
\textit{Institute of Artificial Intelligence} \\
\textit{CISDI Information Technology(Chongqing) Co., Ltd.}\\
Chongqing, 401122, China \\
schemer@sjtu.edu.cn
}
\hfill
\parbox{0.32\textwidth}{
\centering
Shu Liu \\
\textit{National Key Laboratory of Wireless Communications} \\
\textit{University of Electronic Science and Technology of China}\\
Chengdu, China\\
shuliu@uestc.edu.cn
}
\hfill
\parbox{0.32\textwidth}{
\centering
Ningning Wang \\
\textit{School of Computer Science} \\
\textit{Shanghai Jiao Tong University}\\
Shanghai, China \\
wangnning@sjtu.edu.cn
}
}

\maketitle

\begin{abstract}
 Correcting synchronization errors (insertions, deletions, and substitutions) in concatenated coding schemes, typically comprising an inner marker code and an outer Low-Density Parity-Check (LDPC) code, poses a significant challenge. Traditional algebraic decoding fails entirely under sequence misalignment. While recent neural pipelines attempt to resolve this, they typically demand exact channel state information (CSI) for inner synchronization and suffer from error propagation caused by premature syndrome binarization before outer decoding. In this paper, we propose ASDecoder, a data-driven, CSI-agnostic end-to-end neural decoding architecture specifically designed for LDPC codes over insertion-deletion-substitution (IDS) channels. ASDecoder features a CNN-Transformer front-end that implicitly aligns variable-length sequences and fuses multi-read redundancies without requiring explicit CSI. Tightly integrated into this pipeline is a novel \emph{syndrome-free} BiDecoder back-end. By mapping continuous soft features directly into a sparse-masked graph attention network, it strictly avoids intermediate hard decisions and effectively captures topological constraints. Experiments demonstrate that a single universal ASDecoder outperforms the BCJRFormer baseline, providing consistent error correction across highly asymmetric channels. Furthermore, it generalizes robustly to real-world deployment scenarios, including the industrial CCSDS $(128,64)$ QC-LDPC standard and non-linear DNA biological constraints.
\end{abstract}


\section{Introduction}
        Reliable sequence decoding in the presence of synchronization errors remains a major challenge in modern communication and storage systems~\cite{IDSchall,IDSchall2}. 
        Such errors, including insertion, deletion, and substitution (IDS), are typically modeled as an IDS channel. They arise in a wide range of applications such as packet-based communications~\cite{packetCom,wirelessCom}, magnetic storage systems~\cite{magneticstorage}, and DNA data storage~\cite{DNAstorage,DNAstorage2,DNAstorage3}, where retrieving multiple noisy copies (reads) of the same sequence is a common operational characteristic.

        Modern error-correcting codes, particularly Low-Density Parity-Check (LDPC) codes, are widely deployed to approach channel capacity over memoryless channels~\cite{LDPC}. However, their performance degrades over IDS channels. Because LDPC decoding relies on a fixed, sparse bipartite graph (Tanner graph), even a single insertion or deletion shifts the sequence, misaligning the received symbols with their corresponding variable nodes. This misalignment causes a cascade of decoding failures that exceed the code's inherent correction capability.

        To address this, a widely adopted solution is to employ concatenated coding schemes that combine an outer LDPC code with an inner code designed for synchronization, such as marker codes~\cite{marker} or watermark codes~\cite{watermarker}.
        Decoding is typically divided into two stages: an inner decoder resolves synchronizations and outputs soft information, which is then fed into the outer LDPC decoder for error correction. A representative classical approach uses the Bahl--Cocke--Jelinek--Raviv (BCJR) algorithm~\cite{BCJR} for inner alignment, followed by the Belief Propagation (BP) algorithm on the LDPC graph.

        Despite their theoretical foundations, classical IDS decoding methods suffer from practical limitations. First, the computational complexity of BCJR grows exponentially with the number of reads in multi-read scenarios~\cite{BCJRFormer}. Second, BCJR requires precise channel state information (CSI), which is often unavailable or dynamic in practical systems.

        Motivated by the success of deep learning, recent works have proposed neural decoders to replace BCJR~\cite{BIGRU,FBNet,BCJRFormer} and BP~\cite{ECCT,crossMPT}. 
        While these approaches reduce computational complexity and relax certain CSI assumptions, they retain a two-stage paradigm. Specifically, existing neural decoders for linear block codes (e.g., ECCT~\cite{ECCT} and CrossMPT~\cite{crossMPT}) are designed for memoryless channels and require the \emph{syndrome} (calculated via hard decisions) as input. In IDS channels, however, the inner neural synchronizer outputs uncertain, continuous soft features. Forcing intermediate hard decisions on these features to compute the syndrome introduces erroneous priors into the LDPC graph, causing error propagation that degrades outer decoding performance.

        In this work, we propose the \textbf{A}ttention-based \textbf{S}ynchronization \textbf{D}ecoder (\textbf{ASDecoder}), a fully data-driven, CSI-agnostic end-to-end neural decoding architecture specifically designed for LDPC codes over IDS channels. 
        ASDecoder jointly handles inner synchronization and outer error correction by mapping corrupted observations directly to the decoded codeword without requiring explicit CSI or intermediate hard decisions.

        The main contributions of this work are summarized as follows:
        \begin{itemize}
            \item \textbf{CSI-Agnostic Implicit Alignment:} We design a CNN-Transformer front-end that dynamically aligns unaligned sequences. It fuses multi-read redundancies via cross-attention, approaching perfect-CSI performance without requiring explicit channel information.
            
            \item \textbf{Syndrome-Free Structure-Aware Decoding:} Tailored for LDPC codes, we propose the BiDecoder back-end, a sparse-masked graph attention network. By directly ingesting continuous soft features instead of syndromes derived from early hard decisions, the BiDecoder mitigates the error propagation common in existing two-stage neural decoders.
            
            \item \textbf{Universal Model Generalization:} Instead of training separate models for individual error configurations, ASDecoder is trained as a single universal model over a mixture of channel conditions. It maintains stable performance across highly asymmetric channels and generalizes to real-world deployment scenarios, including industrial Quasi-Cyclic LDPC (QC-LDPC) standards and non-linear DNA biological constraints.
        \end{itemize}

    \section{Related Work}
        For IDS channels, most learning-based approaches focus on replacing the BCJR algorithm for inner synchronization decoding.
        The BI-GRU-based decoder proposed in~\cite{BIGRU} employs bidirectional gated recurrent units to estimate LLRs for marker codes, significantly reducing decoding complexity and relaxing the requirement for explicit CSI.
        When combined with convolutional outer codes, the authors further demonstrate that a GRU-based model can be extended to outer decoding.
        However, this approach remains a two-stage pipeline and considers only deletion and substitution errors, without fully addressing general IDS channels.

        \cite{FBNet} extends this line of work by proposing both a model-driven architecture, FBNet, inspired by BCJR, and a purely data-driven BI-GRU-based decoder, FBGRU.
        This framework supports all three types of IDS errors and remains CSI-agnostic, further improving robustness.
        Nevertheless, FBNet focuses exclusively on inner synchronization and relies on classical outer decoders.

        More recently, BCJRFormer~\cite{BCJRFormer} introduced a Transformer-based architecture to approximate BCJR decoding.
        By leveraging attention mechanisms with positional and sequential encodings, BCJRFormer reduces the complexity of multi-read decoding from exponential to quadratic with respect to the number of reads.
        Despite its improved scalability, BCJRFormer requires explicit CSI to construct the drift sliding window for the Transformer input and remains a separated two-stage approach.

        In parallel, learning-based decoding of outer error-correcting codes has attracted significant attention~\cite{rela1,rela2}.
        ECCT~\cite{ECCT} is among the first Transformer-based neural decoders for soft decoding of linear block codes over memoryless channels.
        CrossMPT~\cite{crossMPT} further enhances neural message passing by modeling interactions between variable and check nodes using cross-attention mechanisms.

        However, applying these neural outer decoders directly to IDS channels is problematic.
        Designed for memoryless environments, models like ECCT and CrossMPT fundamentally rely on the \emph{syndrome} as a structural input. 
        For example, when BCJRFormer adapts ECCT for outer decoding, the uncertain soft outputs from the inner synchronizer must undergo premature binarization (intermediate hard decisions) to compute this syndrome. 
        This introduces unreliable priors into the LDPC Tanner graph, causing error propagation and preventing true end-to-end joint optimization across the synchronization and error-correction stages.

    \section{System Model and Problem Formulation}\label{sec:sys_mod}
\begin{figure}[!htbp]
            \centering
        \includegraphics[width=\linewidth]{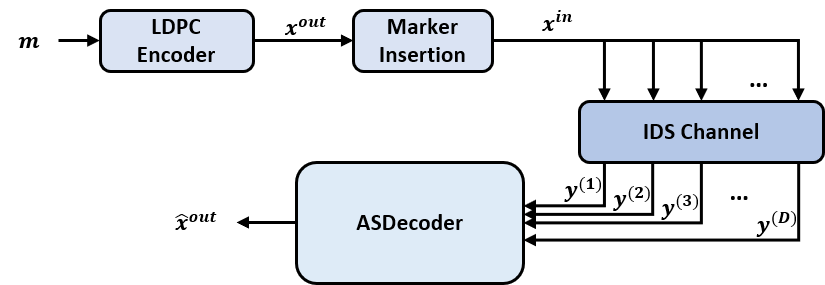}
            \caption{System model of a concatenated coding scheme over an IDS channel.}
            \label{fig:system_model}
        \end{figure}

    We investigate a binary concatenated coding system transmitted over insertion, deletion and substitution (IDS) channels, whose overall architecture is depicted in Fig.~\ref{fig:system_model}.
    \iffalse
     We consider a binary concatenated coding system operating over an IDS channel, as illustrated in Fig.~\ref{fig:system_model}. 
\fi
       
        At the transmitter, a $k$-bit message $\mathbf{m}\in\{0,1\}^k$ is encoded into an $n$-bit outer codeword $\mathbf{x}^{\text{out}}\in\{0,1\}^n$ using an $(n,k)$ LDPC code. This code is characterized by a sparse parity-check matrix $\mathbf{H} \in \{0,1\}^{(n-k) \times n}$, which will be explicitly leveraged by our neural decoder. To facilitate synchronization, a predefined $n_m$-bit marker $\mathbf{s}_m\in\{0,1\}^{n_m}$ is inserted periodically into $\mathbf{x}^{\text{out}}$ every $N_m$ bits, resulting in an inner codeword $\mathbf{x}^{\text{in}}\in\{0,1\}^N$ of length
        \[
        N = n + \left\lfloor \frac{n}{N_m} \right\rfloor n_m.
        \]

\begin{figure}[!htbp]
            \centering
            \includegraphics[width=0.9\linewidth]{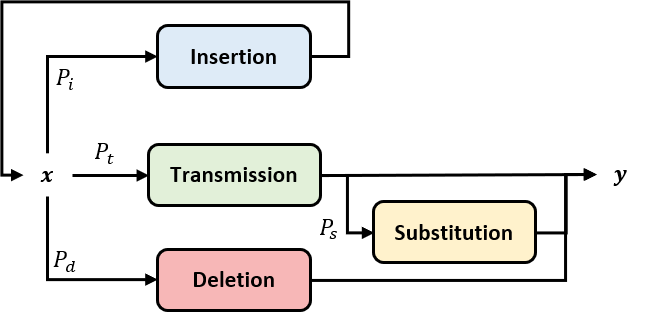}
            \caption{The IDS channel model showing the three possible events for each input symbol}
            \label{fig:ids}
        \end{figure}

        The codeword $\mathbf{x}^{\text{in}}$ is then transmitted through the IDS channel. The IDS channel is characterized by three parameters: insertion probability $P_i$, deletion probability $P_d$, and substitution probability $P_s$~\cite{FBNet}.
        As shown in Fig.~\ref{fig:ids}, for each input bit $x$, one of three mutually exclusive events happens: i) a random bit is inserted before $x$ with probability $P_i$ and $x$ remains untransmitted; ii) $x$ is deleted with probability $P_d$; iii) $x$ is transmitted with probability $P_t=1-P_i-P_d$, but flipped with probability $P_s$.

        In practical applications such as DNA data storage, processes like PCR amplification and sequencing naturally generate a cluster of noisy reads for the same transmitted sequence. Accordingly, the receiver observes a set of $D$ received sequences corresponding to the same transmitted codeword
        \[
        \mathbf{Y} = \left\{ \mathbf{y}^{(d)} \right\}_{d=1}^{D},
        \]
        where each $\mathbf{y}^{(d)}$ may have a different length due to insertion and deletion events. The decoding objective is to recover the original outer codeword $\mathbf{x}^{\text{out}}$ from $\mathbf{Y}$, without requiring prior knowledge of the channel state information (CSI) (i.e., $P_i, P_d, P_s$) in Fig.~\ref{fig:ids}.
        
    \section{End-to-End Neural Decoding over IDS Channels}\label{sec:method}
\begin{figure}[!htbp]
            \centering
    \includegraphics[width=1\linewidth]{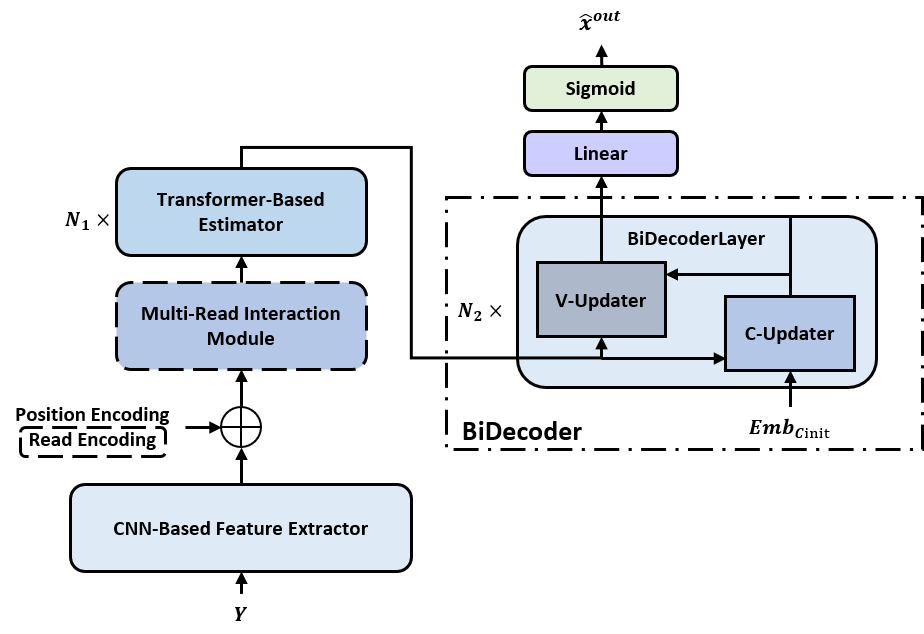}
            \caption{Overall architecture of ASDecoder.}
            \label{fig:overview}
        \end{figure}

        In this section, we present the details of the proposed ASDecoder for IDS channels under the system model described in Section~\ref{sec:sys_mod}. An overview of the architecture is shown in Fig.~\ref{fig:overview}.
        A central design objective of ASDecoder is to flexibly support multiple received sequences corresponding to the same transmitted codeword.
        To this end, the decoder provides two operating modes for multi-read processing:
        \begin{itemize}
            \item \textit{Channel-stacked (CS) mode (default):} multiple reads are stacked along the channel dimension and processed jointly by the CNN front-end.
            \item \textit{Alignment-enhanced (AE) mode:} each read is explicitly preserved and processed independently by the CNN before being fed into the estimator.
        \end{itemize}
        Both modes share the same high-level architecture and training procedure, differing only in how multi-read information is fused.
        This design enables a flexible trade-off between decoding performance and computational efficiency.

        \subsection{CNN-Based Feature Extractor}
            The CNN-based feature extractor is responsible for extracting local, CSI-agnostic representations from received sequences corrupted by insertion, deletion, and substitution errors.
            After padding, the received observations are represented as
            $\mathbf{Y} \in \mathbb{R}^{B \times D \times T}$,
            where $B$ denotes the batch size, $D$ the number of reads, and $T$ the padded sequence length.
            For both operating modes, convolutional blocks with residual connections and group normalization are employed throughout the CNN to stabilize training.

            \paragraph{Channel-stacked (CS) mode.}
            All reads are concatenated along the channel dimension and processed jointly by the CNN as the following procedure:
            \[
            \mathbf{Y} \in \mathbb{R}^{B \times D \times T}
            \;\xrightarrow{\text{Conv1D}}\;
            \mathbf{H} \in \mathbb{R}^{B \times d_{\text{model}} \times T}.
            \]
            This provides an efficient approach to merge multiple reads. In this mode, the first convolutional layer extracts and fuses all $D$ reads to produce one single feature representation. While this early fusion may introduce some information loss, the computational complexity is significantly reduced since only the first layer scales with $D$, making the overall feature extraction complexity essentially $\mathcal{O}(1)$ with respect to the number of reads. Following the CNN blocks, the tensor is permuted to $\mathbb{R}^{B \times T \times d_{\text{model}}}$, and standard sinusoidal positional encodings are added along the time axis to provide location information.

            \paragraph{Alignment-enhanced (AE) mode.}
            To prevent early information bottlenecking, each read is explicitly preserved and processed independently.
            The input tensor is first reshaped as
            \[
            \mathbf{Y} \in \mathbb{R}^{B \times D \times T}
            \;\longrightarrow\;
            \mathbf{Y}' \in \mathbb{R}^{(B \cdot D) \times 1 \times T},
            \]
            and passed through the CNN:
            \[
            \mathbf{Y}' \xrightarrow{\text{Conv1D}}
            \mathbf{H} \in \mathbb{R}^{(B \cdot D) \times d_{\text{model}} \times T}.
            \]
            The resulting features are rearranged into
            \[
            \mathbf{F} \in \mathbb{R}^{B \times T \times D \times d_{\text{model}}}.
            \]
            This mode sacrifices computational efficiency (as the CNN complexity scales linearly as $\mathcal{O}(D)$) but explicitly preserves inter-read distinctions, providing richer context for multi-read processing.

            To exploit these features, two sets of learnable positional embeddings are introduced, corresponding to the read index and the temporal position within each read, respectively. After adding these embeddings, the tensor is reshaped to $\mathbb{R}^{(B \cdot T) \times D \times d_{\text{model}}}$ to facilitate an explicit inter-read attention mechanism. For each batch index $b$ and implicit time step $t$, the slice $\mathbf{F}_{b,t} \in \mathbb{R}^{D \times d_{\text{model}}}$ is processed by a multi-head self-attention layer
            \[
            \mathbf{Z}_{b,t} = \mathrm{MHA}\!\left( \mathbf{F}_{b,t}, \mathbf{F}_{b,t}, \mathbf{F}_{b,t} \right).
            \]
            This operation explicitly aggregates complementary consensus information across reads at the same implicit temporal index. The interaction is strictly restricted to the read dimension, capturing cross-sequence correlations without performing premature alignment or drift estimation. Finally, the token sequence is rearranged and flattened into a unified representation $\mathbf{M} \in \mathbb{R}^{B \times (D \cdot T) \times d_{\text{model}}}$ to serve as the input memory for the subsequent Estimator.       
        
        \subsection{Transformer-Based Estimator}
            The Transformer-based estimator is designed to align the variable-length memory representations extracted from either a single fused read or multiple independent reads into a fixed length-$n$ feature sequence corresponding to the outer codeword.

            While both our estimator and recent methods like BCJRFormer utilize Transformer architectures for synchronization, they differ significantly in their alignment mechanisms. BCJRFormer relies on explicit pre-alignment: it uses perfect CSI to determine the lower and upper bounds of synchronization drift and constructs fixed-size sliding windows, forming an input of length $n$ (or $n \cdot D$ for multi-read scenarios). This pre-aligned sequence is then processed by a self-attention encoder, followed by a straightforward averaging operation across the $D$ sequences to fuse multi-read features.

            In contrast, ASDecoder performs implicit alignment without requiring explicit CSI or pre-calculated drift windows. We employ a query-based cross-attention architecture, where $n$ learnable target queries directly attend to the unaligned, variable-length memory sequence extracted by the CNN. This mechanism allows the model to dynamically learn the alignment and extract a unified length-$n$ representation directly. Furthermore, by avoiding self-attention over the concatenated sequence of length $n \cdot D$, our estimator reduces the alignment complexity from $\mathcal{O}((nD)^2)$ in BCJRFormer to $\mathcal{O}(n^2 D)$ in multi-read scenarios.
        
        \subsection{Structure-Aware BiDecoder}
            The final decoding stage is performed by a structure-aware \emph{BiDecoder}, which enforces the algebraic constraints of the outer LDPC code via neural message passing. Let $\mathbf{H} \in \{0,1\}^{(n-k) \times n}$ denote the sparse parity-check matrix of the outer code.

            Unlike existing neural decoders such as ECCT and CrossMPT, which are designed for memoryless channels and typically concatenate continuous soft magnitudes with hard-decoded syndromes, the BiDecoder is strictly \textbf{syndrome-free}. In IDS channels, sequence length variations make it impossible to compute valid syndromes prior to alignment without resorting to premature hard decisions. To prevent the potential error propagation caused by such hard decision, our BiDecoder directly ingests the continuous, length-$n$ soft features output by the estimator as the initial embeddings of the variable nodes. To force the network to learn parity-check constraints purely from the graph topology rather than relying on unreliable syndromes computed from early hard decisions, the check node embeddings are initialized as all-zero vectors of length $(n-k)$.

            The BiDecoder iteratively refines these embeddings using a stack of \emph{BiDecoderLayers}. Based on the Transformer decoder architecture, each layer sequentially alternates between node-level self-attention and bipartite cross-attention. Specifically, it consists of:
            (i) a check node update, where check nodes first perform dense self-attention to capture global parity correlations, and then attend only to their neighboring variable nodes using $\mathbf{H}^\top$ as the cross-attention mask.
            (ii) a variable node update, where variable nodes similarly perform self-attention, followed by cross-attention to their connected check nodes using $\mathbf{H}$ as the mask.

            In this hybrid design, the dense self-attention modules allow for global information routing among nodes of the same type with a standard $\mathcal{O}(n^2)$ complexity, the structural prior enforced by the sparse mask ensures that the bipartite message passing occurs exclusively along valid edges of the factor graph. Standard dense cross-attention between $n$ variable nodes and $(n-k)$ check nodes would incur a quadratic computational complexity of $\mathcal{O}(n(n-k))$. By explicitly masking invalid connections, our method strictly reduces this inter-node cross-attention complexity to $\mathcal{O}(|E|)$, where $|E|$ is the number of edges in the Tanner graph. Because LDPC matrices are inherently sparse (e.g., $|E| = d_v \cdot n$ for a regular LDPC code with a constant variable degree $d_v$), the edge count scales linearly with the code length $n$. This linear scaling mitigates the quadratic cross-attention bottleneck, allowing the BiDecoder to scale efficiently to longer codes.

            To reduce the number of parameters, the weights of the BiDecoderLayer can be shared across iterations, effectively unrolling a neural message-passing process analogous to Belief Propagation (BP). After a fixed number of iterations, the refined variable node embeddings are passed through a linear layer to produce bit-level logits (without sigmoid).
        
        \subsection{End-to-End Training}
            The entire ASDecoder pipeline, which comprises the CNN feature extractor, the Transformer-based estimator, and the structure-aware BiDecoder, is optimized end-to-end using supervised learning.

            Given the transmitted outer codeword $\mathbf{x}^{\text{out}} \in \{0,1\}^n$, the model directly outputs the 1D bit-level logits $\hat{\mathbf{y}}$. We employ the standard binary cross-entropy (BCE) loss:
            \[
            \mathcal{L}_{\mathrm{BCE}}
            = - \frac{1}{n} \sum_{i=1}^{n}
            \big[
            x^{\text{out}}_i \log \sigma(\hat{y}_i)
            + (1-x^{\text{out}}_i)\log (1-\sigma(\hat{y}_i))
            \big],
            \]
            where $\sigma(\cdot)$ denotes the sigmoid function.

            Notably, the architecture requires no intermediate supervision (such as explicit alignment targets or drift estimation loss). By backpropagating the BCE loss through all components jointly, the network implicitly learns the synchronization and feature alignment that serve the algebraic constraints of the outer LDPC code.
            
    % ablation part
    % more parameters of ldpc code
    % DNA part rewrite
    \section{Experimental Results}\label{sec:experiments}
        In this section, we evaluate the proposed ASDecoder under various IDS channel conditions.
        We compare against a representative learning-based baseline: BCJRFormer~\cite{BCJRFormer}. 
        Classical BCJR+BP is omitted from the baselines, as prior work~\cite{BCJRFormer} has established the advantages of neural decoders over classical algorithms in severe IDS settings.
        
        \subsection{Experimental Setup}
            Our experimental setup employs rate-$1/2$ LDPC codes with randomly generated non-zero codewords.
            We measure the performance using bit error rate (BER), which is the ratio of wrong bits of the decoded codeword.
            To evaluate the generalization capability of ASDecoder, we consider three distinct LDPC codes: two standard (3,6)-regular LDPC codes commonly used in prior work, $(96,48)$ and $(204,102)$, and a highly structured $(128,64)$ Quasi-Cyclic LDPC (QC-LDPC) code compliant with the CCSDS standard. 
            % The CCSDS code features a lifting factor of $Z=16$ and a sum-of-circulants structure, serving to validate the model's applicability to industrial-grade factor graphs.

            Unless otherwise stated, all models are trained with binary cross-entropy loss using the Adam optimizer.
            The learning rate is set to $5\times 10^{-4}$ with cosine decay and a $10\%$ warm-up of total steps. All models are trained for $160{,}000$ steps with a batch size of $256$. All training and inference runtime evaluations are conducted on an NVIDIA RTX 3090 GPU to ensure fair computational comparisons.

            Marker-based inner codes are used in all experiments.
            The marker pattern and insertion interval are chosen to match those used by the corresponding baseline methods. For the $(204,102)$ code, we use the marker pattern $\mathbf{s}_m=001$ with an insertion interval of $N_m=7$. For the $(96,48)$ code, we use the same marker with $N_m=6$.
            
            For the ASDecoder, the CNN front-end consists of two 1D convolutional layers with kernel sizes $7$ and $5$, respectively.

            When comparing front-end alignment performance with BCJRFormer (Section~\ref{sec:exp_frontend}), we strictly align the model capacity of our Transformer-based estimator to that of BCJRFormer. Specifically, we set the hidden dimension $d_{\text{model}}=96$, feedforward dimension $d_{\text{ff}}=384$, and use $6$ layers.
            The total number of trainable parameters is approximately $1.0$M.

            For all other experiments, we use a slightly larger configuration with $d_{\text{model}}=128$ and $d_{\text{ff}}=512$. The Transformer-based estimator uses $3$ layers, and the BiDecoder repeats a single BiDecoderLayer $6$ times. This configuration has approximately $1.58$M parameters.

            \textbf{Universal Model Training Strategy:} Unlike BCJRFormer, which trains a separate independent model for each specific channel condition, we train ASDecoder as a single \emph{universal} model to evaluate its robustness under dynamic CSI uncertainty. During training, we construct a highly mixed dataset comprising 100 distinct channel configurations ($P_i \in [0.005, 0.05]$ and $P_d \in [0.005, 0.05]$ with a step size of $0.005$). At inference time, this single model is evaluated across all conditions. For visual clarity and concise presentation, both the front-end performance (Section~\ref{sec:exp_frontend}) and end-to-end results (Section~\ref{sec:exp_e2e}) are reported along the symmetric error diagonal ($P_i=P_d$).

        \subsection{Front-End Alignment Performance}\label{sec:exp_frontend}
            We first evaluate the \emph{front-end alignment capability} of ASDecoder and compare it with BCJRFormer.
            To isolate the alignment performance from the algebraic decoding gains, the memory sequence output by the Transformer-based estimator is directly mapped to bit-level logits using a simple linear layer, bypassing the structure-aware BiDecoder. This setting allows us to focus exclusively on the ability of the neural front-end to handle severe synchronization errors.
        
\begin{figure}[!htbp]
                \centering
                \includegraphics[width=1.05\linewidth]{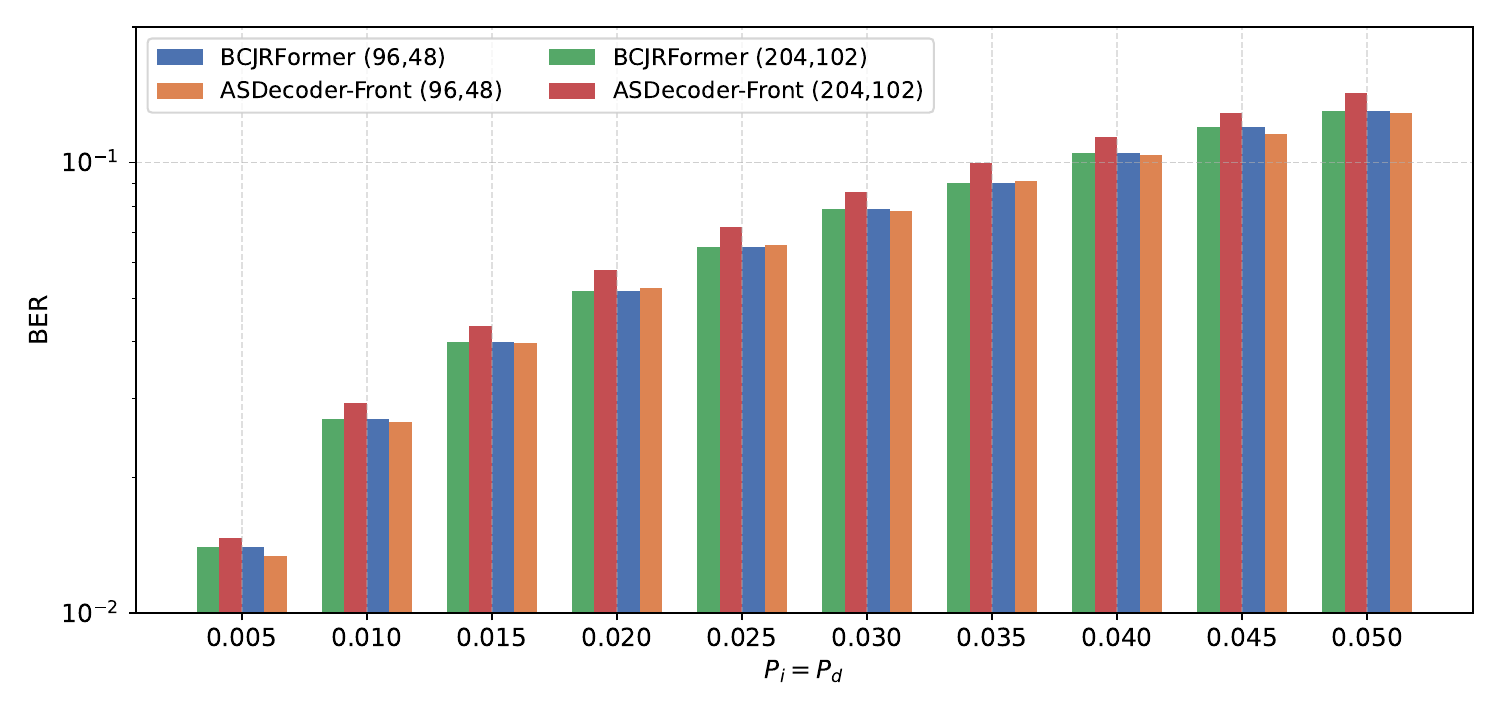}
                \caption{Single-read front-end BER comparison ($D=1$, $P_s=0$) on $(96,48)$ and $(204,102)$ LDPC codes.}
                \label{fig:exp1_single}
            \end{figure}

            \paragraph{Single-read decoding ($D=1$).}
            Figure~\ref{fig:exp1_single} shows the single-read alignment results with $P_s=0$. In this setting, only the CS variant of ASDecoder is evaluated.

            For the shorter $(96,48)$ code, ASDecoder closely matches BCJRFormer across all choices of $P_i=P_d$. For the longer $(204,102)$ code, ASDecoder exhibits a slightly higher BER. This gap is expected: BCJRFormer leverages perfect CSI to explicitly compute exact drift bounds and construct pre-aligned sliding windows, an advantage that becomes more pronounced as sequence length increases. In contrast, our ASDecoder operates without explicit CSI, dynamically learning implicit alignments. Given the absence of perfect CSI, ASDecoder maintains competitive performance relying solely on the CNN-Transformer feature extraction.
            
\begin{figure}[!htbp]
                \centering
                \includegraphics[width=1.05\linewidth]{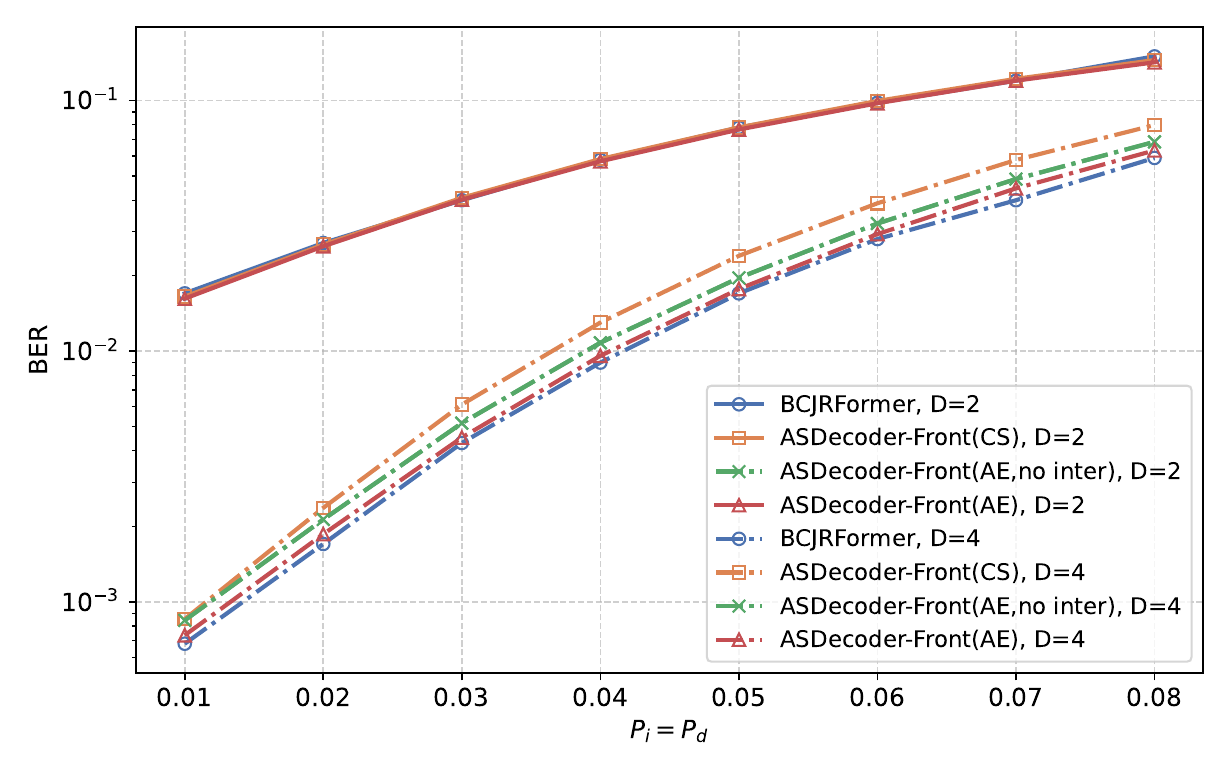}
                \caption{Multi-read front-end BER comparison on the $(96,48)$ code with $P_s=0.012$ and $D\in\{2,4\}$.}
                \label{fig:exp1_multi}
            \end{figure}

            \paragraph{Multi-read decoding ($D>1$).}
            To demonstrate how ASDecoder compensates for the lack of CSI in practical scenarios, Figure~\ref{fig:exp1_multi} reports multi-read front-end performance on the $(96,48)$ code with $P_s=0.012$. We compare BCJRFormer with both the CS and AE variants of ASDecoder. Additionally, for the AE variant, we conduct an ablation to isolate the effect of the inter-read attention mechanism.
    
            When $D=2$, the performance of all variants is nearly identical, indicating that simple channel stacking (CS mode) is sufficient when only minimal redundancy is available. However, as sequence redundancy increases at $D=4$, the early fusion bottleneck of the CS variant causes noticeable performance degradation. In contrast, the AE variant explicitly preserves inter-read distinctions.

            The ablation reveals that the full AE variant (with explicit inter-read attention mechanism) outperforms its non-interacting counterpart, approaching the performance of the perfect-CSI BCJRFormer. This indicates that allowing independent reads to attend to one another via cross-attention helps build a dynamic synchronization consensus, substituting the need for explicit channel drift knowledge.
            
                    \subsection{End-to-End Performance and Ablation Study}\label{sec:exp_e2e}
            We evaluate the end-to-end performance and conduct an ablation study against the strongest two-stage baseline, BCJRFormer+ECCT. We use the $(96,48)$ code evaluating $512{,}000$ codewords.

\begin{figure}[!htbp]
                \centering
                \includegraphics[width=\linewidth]{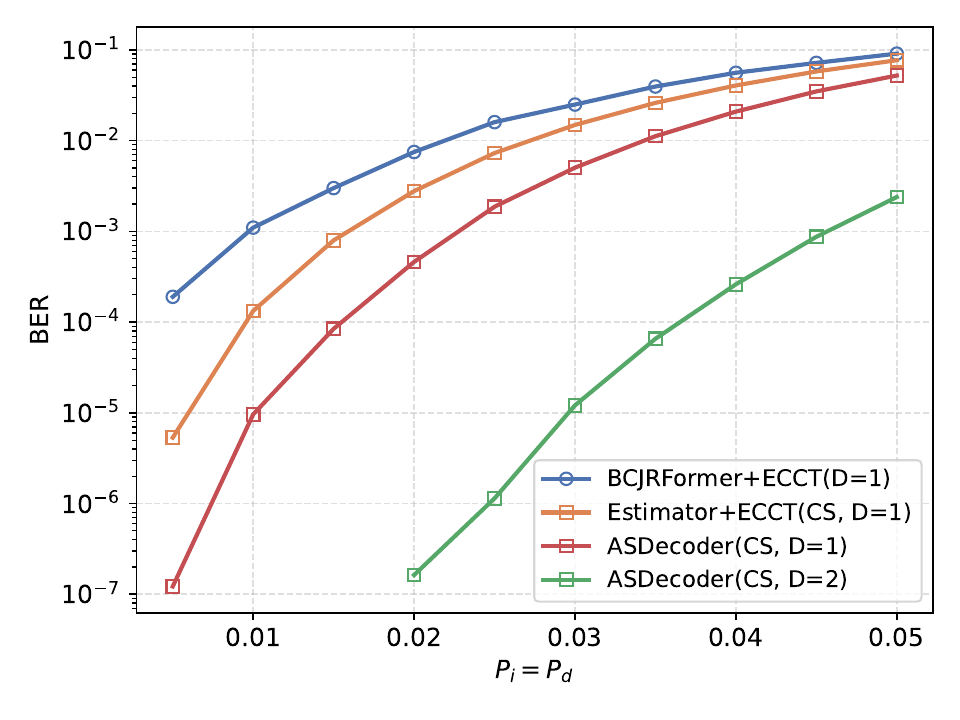}
                \caption{End-to-end decoding performance and ablation study on the $(96,48)$ LDPC code with $P_s=0$.}
                \label{fig:exp2_e2e}
            \end{figure}

            Figure~\ref{fig:exp2_e2e} isolates the performance gains in single-read ($D=1$) scenarios. To validate joint training, we test \emph{Estimator+ECCT} (trained end-to-end with zero-syndrome initialization). At an error rate of $P_i=P_d=0.03$, the BER drops from $2.50 \times 10^{-2}$ (baseline) to $1.49 \times 10^{-2}$, proving that avoiding intermediate hard decisions for syndrome computation prevents error propagation. Furthermore, replacing ECCT with our syndrome-free \emph{BiDecoder} (forming ASDecoder) further reduces the BER to $5.04 \times 10^{-3}$. This indicates that the sparse-masked architecture effectively extracts topological constraints from unaligned features.
            Additionally, an unshared-weight ASDecoder variant yields an almost identical BER (e.g., $5.05 \times 10^{-3}$ at $P_i=0.03$) but significantly increases the parameter count, supporting the use of the weight-sharing strategy. Finally, utilizing one additional read ($D=2$) further reduces the BER by orders of magnitude (e.g., $1.21 \times 10^{-5}$ at $P_i=0.03$).

        \subsection{Practical Applications}\label{sec:exp_dna}
            We further evaluate ASDecoder in two real-world scenarios: industrial communication standards and biological data storage.
    
            \paragraph{Generalization to Industrial QC-LDPC Codes.}
            We evaluate structural generalization on the $(128, 64)$ QC-LDPC code compliant with the CCSDS standard. This industrial-grade code features a lifting factor of $Z=16$ and a complex sum-of-circulants parity-check matrix widely used in deep-space communications. 
            
            Table~\ref{tab:ccsds_ber} details the end-to-end decoding performance of the universal ASDecoder across a comprehensive grid of asymmetric insertion and deletion probabilities ($P_s=0$). The single trained model provides consistent error correction across all combinations. It reaches a BER of $7.05 \times 10^{-6}$ at $P_i=P_d=0.01$, and maintains stable decoding performance under severely imbalanced channel conditions (e.g., yielding $9.73 \times 10^{-3}$ at $P_i=0.01, P_d=0.05$). 
            
            \begin{table}[h]
                \centering
                \caption{End-to-End decoding performance on the $(128,64)$ CCSDS code with $P_s=0$.}
                \label{tab:ccsds_ber}
                \resizebox{1.05\linewidth}{!}{
                \begin{tabular}{lccccc}
                    \toprule
                    \textbf{$P_i \setminus P_d$} & 0.01 & 0.02 & 0.03 & 0.04 & 0.05 \\
                    \midrule
                    0.01 & $7.05 \times 10^{-6}$ & $1.23 \times 10^{-4}$ & $9.05 \times 10^{-4}$ & $3.58 \times 10^{-3}$ & $9.73 \times 10^{-3}$ \\
                    0.02 & $7.59 \times 10^{-5}$ & $7.71 \times 10^{-4}$ & $3.43 \times 10^{-3}$ & $9.88 \times 10^{-3}$ & $2.15 \times 10^{-2}$ \\
                    0.03 & $3.83 \times 10^{-4}$ & $2.50 \times 10^{-3}$ & $8.55 \times 10^{-3}$ & $2.01 \times 10^{-2}$ & $3.77 \times 10^{-2}$ \\
                    0.04 & $1.23 \times 10^{-3}$ & $6.17 \times 10^{-3}$ & $1.69 \times 10^{-2}$ & $3.40 \times 10^{-2}$ & $5.63 \times 10^{-2}$ \\
                    0.05 & $3.29 \times 10^{-3}$ & $1.24 \times 10^{-2}$ & $2.86 \times 10^{-2}$ & $5.09 \times 10^{-2}$ & $7.70 \times 10^{-2}$ \\
                    \bottomrule
                \end{tabular}
                }
            \end{table}

            \paragraph{Decoding under DNA Biological Constraints.}
            We apply ASDecoder to DNA data storage, which is naturally modeled as an IDS channel~\cite{DNAaschannel}. To ensure biochemical stability, valid DNA sequences typically need to satisfy multiple biological requirements. In this work, we focus on two primary constraints: global GC balance $$\left|\text{wt}_{\text{GC}}({\bf x}) - n/2\right| \leq \epsilon n, \epsilon>0$$ and a maximum run-length limit ($\ell$) which restricts the maximum number of consecutive identical nucleotides. Following common configurations in previous DNA coding research~\cite{GCRLdef}, we set $\epsilon=0.1$ and $\ell=5$.
            
            We map the binary $(204,102)$ codewords to DNA sequences (00$\rightarrow$A, 01$\rightarrow$C, 10$\rightarrow$G, 11$\rightarrow$T) and dynamically filter the dataset using marker $\mathbf{s}_m=001$ at intervals of $N_m=6$, achieving a $>99\%$ constraint success rate.
            
\begin{figure}[!htbp]
                \centering
                \includegraphics[width=\linewidth]{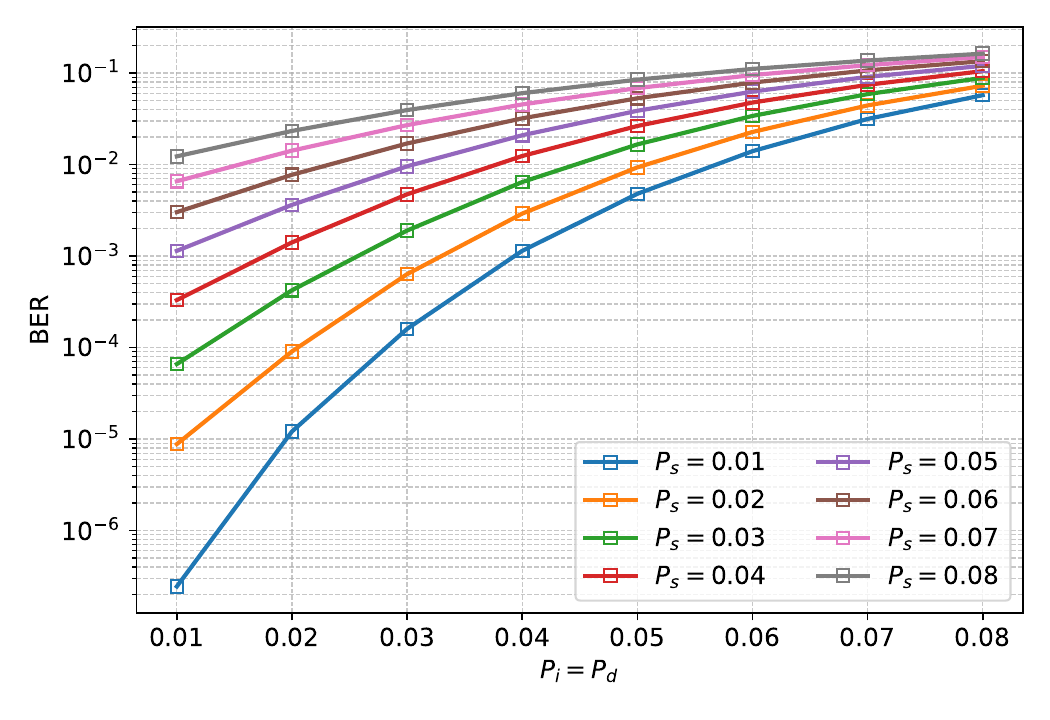}
                \caption{End-to-end decoding performance under biological constraints on the $(204,102)$ LDPC code.}
                \label{fig:exp4_bio_constraints}
            \end{figure}

            The data-driven ASDecoder processes these constrained sequences without algorithmic adjustments. Trained as a universal model over 2-bit symbol errors ($P_i, P_d, P_s \in [0.01, 0.08]$), it maintains stable decoding performance across varying error conditions, as illustrated in Figure~\ref{fig:exp4_bio_constraints}. This demonstrates the practical applicability of the proposed architecture for biological storage systems.

    \section{Conclusion}
        In this paper, we proposed \textbf{ASDecoder}, a data-driven, CSI-agnostic end-to-end neural decoding architecture specifically designed for LDPC codes over insertion-deletion-substitution (IDS) channels. 
        Unlike traditional two-stage decoding pipelines that rely on exact CSI and intermediate hard decisions, ASDecoder maps corrupted sequences directly to the decoded codeword.

        The proposed architecture integrates a CNN-Transformer front-end for implicit alignment and multi-read fusion, paired with a novel \emph{syndrome-free} BiDecoder back-end. 
        By ingesting continuous soft features directly into a sparse-masked graph attention network, this design avoids the error propagation caused by premature binarization and effectively captures the topological constraints of the outer code.
                
        Experimental results demonstrate that ASDecoder outperforms the BCJRFormer baseline. Trained as a single universal model, it provides consistent error correction across highly asymmetric channel conditions. Furthermore, the framework generalizes robustly to real-world deployment scenarios, successfully decoding both the industrial CCSDS $(128,64)$ QC-LDPC standard and sequences subject to non-linear DNA biological constraints.
        
        Overall, ASDecoder provides a unified and scalable solution for synchronization errors, highlighting the effectiveness of end-to-end, structure-aware neural decoding for complex channels.

        While ASDecoder demonstrates strong generalization and robustness, several avenues remain for future exploration. First, the current framework relies on inner marker codes to provide synchronization anchors. Extending this architecture to fully \emph{markerless} synchronization paradigms would further improve the overall coding rate. Second, in practical DNA data storage systems, multi-read retrieval often suffers from sequence clustering errors, such as the accidental inclusion of reads from different codewords or missing reads. Investigating the model's resilience against such out-of-cluster noise is an important next step for real-world deployment. Third, although the BiDecoder is tailored for the sparse Tanner graphs of LDPC codes, adapting the sparse-masked mechanism for other prominent code families, such as Polar or BCH codes, remains to be explored. Finally, further optimizing the attention mechanisms to maintain efficient scaling with respect to massive code lengths and high read counts constitutes a promising direction for future research.

%\bibliography{aaai2027}

\begin{thebibliography}{00}
\bibitem{BCJR}
L.~R. Bahl, J. Cocke, F. Jelinek, and J. Raviv, ``Optimal decoding of linear codes for minimizing symbol error rate (Corresp.),'' \emph{IEEE Trans. Inf. Theory}, vol.~20, no.~2, pp.~284--287, 1974.
\bibitem{DNAstorage3}
C. Bancroft, B. Bowler, B. Bloom, and C.~T. Clelland, ``Long-term storage of information in DNA,'' \emph{Science}, vol.~293, no.~5536, pp.~1763--1765, 2001.
\bibitem{packetCom}
A. Checko, A.~C. Juul, H.~L. Christiansen, and M.~S. Berger, ``Synchronization challenges in packet-based CloudRAN fronthaul for mobile networks,'' in \emph{IEEE International Conference on Communication (ICC)}, London, United Kingdom, Jun. 2015, pp.~2721--2726.
\bibitem{wirelessCom}
W. Chen and Y. Liu, ``Efficient transmission schemes for correcting insertions/deletions in DPPM,'' in \emph{IEEE International Conference on Communications (ICC)}, Kuala Lumpur, Malaysia, May 2016, pp.~1--6.
\bibitem{ECCT}
Y. Choukroun and L. Wolf, ``Error Correction Code Transformer,'' in \emph{Advances in Neural Information Processing Systems (NeurIPS)}, vol.~35, New Orleans, LA, USA, Nov. 2022.
\bibitem{watermarker}
M.~C. Davey and D.~J.~C. MacKay, ``Reliable communication over channels with insertions, deletions, and substitutions,'' \emph{IEEE Trans. Inf. Theory}, vol.~47, no.~2, pp.~687--698, 2001.
\bibitem{LDPC}
R.~G. Gallager, ``Low-density parity-check codes,'' \emph{IRE Trans. Inf. Theory}, vol.~8, no.~1, pp.~21--28, 1962.
\bibitem{DNAaschannel}
R. Heckel, G. Mikutis, and R.~N. Grass, ``A Characterization of the DNA Data Storage Channel,'' CoRR, abs/1803.03322, 2018.
\bibitem{BIGRU}
E.~U. Kargi and T.~M. Duman, ``A Deep Learning Based Decoder for Concatenated Coding Over Deletion Channels,'' in \emph{IEEE International Conference on Communications (ICC)}, Denver, CO, USA, Jun. 2024, pp.~2797--2802.
\bibitem{rela1}
H. Kim, Y. Jiang, S. Kannan, S. Oh, and P. Viswanath, ``Deepcode: Feedback Codes via Deep Learning,'' in \emph{Advances in Neural Information Processing Systems (NeurIPS)}, vol.~31, Montréal, Canada, Dec. 2018, pp.~9458--9468.
\bibitem{GCRLdef}
S. Liu, K. Wu, and Y. Zhang, ``Code Constructions for DNA-Based Data Storage,'' in \emph{IEEE Global Communications Conference (GLOBECOM)}, 2025, pp.~1202--1207.
\bibitem{FBNet}
G. Ma, X. Jiao, J. Mu, H. Han, and Y. Yang, ``Deep learning-based detection for marker codes over insertion and deletion channels,'' \emph{IEEE Transactions on Communications}, vol.~72, no.~10, pp.~5945--5959, 2024.
\bibitem{IDSchall2}
H. Mercier, V.~K. Bhargava, and V. Tarokh, ``A survey of error-correcting codes for channels with symbol synchronization errors,'' \emph{IEEE Commun. Surv. Tutorials}, vol.~12, no.~1, pp.~87--96, 2010.
\bibitem{rela2}
E. Nachmani, Y. Be'ery, and D. Burshtein, ``Learning to decode linear codes using deep learning,'' in \emph{Allerton Conference on Communication, Control, and Computing}, Monticello, IL, USA, Sep. 2016, pp.~341--346.
\bibitem{crossMPT}
S. Park, H. Kwak, S. Kim, Y. Kim, and J. No, ``CrossMPT: Cross‑attention Message‑passing Transformer for Error Correcting Codes,'' in \emph{International Conference on Learning Representations (ICLR)}, Singapore, Apr. 2025.
\bibitem{marker}
E.~A. Ratzer, ``Marker codes for channels with insertions and deletions,'' \emph{Ann. des Télécommunications}, vol.~60, no.~1‑2, pp.~29--44, 2005.
\bibitem{DNAstorage2}
I. Shomorony and R. Heckel, ``Information‑Theoretic Foundations of DNA Data Storage,'' \emph{Found. Trends Commun. Inf. Theory}, vol.~19, no.~1, pp.~1--106, 2022.
\bibitem{BCJRFormer}
J. Streit, F. Weindel, and R. Heckel, ``Transformer‑Based Decoding in Concatenated Coding Schemes Under Synchronization Errors,'' in \emph{IEEE International Symposium on Information Theory (ISIT)}, Ann Arbor, MI, USA, Jun. 2025, pp.~1--6.
\bibitem{IDSchall}
I. Tinnirello, D. Croce, N. Galioto, D. Garlisi, and F. Giuliano, ``Cross‑Technology WiFi/ZigBee Communications: Dealing With Channel Insertions and Deletions,'' \emph{IEEE Commun. Lett.}, vol.~20, no.~11, pp.~2300--2303, 2016.
\bibitem{magneticstorage}
R.~L. White, R. Newt, and R.~F.~W. Pease, ``Patterned media: a viable route to 50 Gbit/in/sup 2/and up for magnetic recording?,'' \emph{IEEE Transactions on Magnetics}, vol.~33, no.~1, pp.~990--995, 1997.
\bibitem{{DNAstorage}
S.~M.~H. T. Yazdi, H.~M. Kiah, E.~R. Garcia, J. Ma, H. Zhao, and O. Milenkovic, ``DNA‑Based Storage: Trends and Methods,'' \emph{IEEE Trans. Mol. Biol. Multi Scale Commun.}, vol.~1, no.~3, pp.~230--248, 2015.
\end{thebibliography}


\iffalse
\section*{References}
\fi

\iffalse

\fi

\end{document}